\documentclass[article]{emulateapj}

\usepackage{times}

\received{2011 June 10} 
\accepted{2011 August 3}
\shorttitle{Wavelength dependence of CO ice photodesorption}
\shortauthors{Fayolle et al.}

\begin{document}

\title{CO ice photodesorption: a wavelength-dependent study}

%%%%%%%%%% Author data %%%%%%%%%%%%%%%%%

 \author{Edith C. Fayolle\altaffilmark{1},  Mathieu Bertin\altaffilmark{2}, Claire Romanzin\altaffilmark{2,3}, Xavier Michaut\altaffilmark{2}, Karin I. \"Oberg\altaffilmark{4}, Harold Linnartz\altaffilmark{1}, and Jean-Hugues Fillion\altaffilmark{2}}

\altaffiltext{1}{Sackler Laboratory for Astrophysics, Leiden Observatory, Leiden University, PO Box 9513, 2300 RA Leiden, The Netherlands}
\altaffiltext{2}{Laboratoire de Physique Mol\'{e}culaire pour l'Atmosph\`{e}re et l'Astrophysique, Universit\'{e} Pierre et Marie Curie-Paris 6, CNRS UMR7092, 75005 Paris, France}
\altaffiltext{3}{present address: Laboratoire de Chimie Physique, Universit\'{e} Paris Sud 11, CNRS UMR 8000, 91405 Orsay, France}
 \altaffiltext{4}{Harvard-Smithsonian Center for Astrophysics, 60 Garden Street, Cambridge, MA 02138, USA}

%%%%%%%%% Abstract %%%%%%%%%%%%%%%%
  
\begin{abstract}
UV-induced photodesorption of ice is a non-thermal evaporation process that can explain the presence of cold molecular gas in a range of interstellar regions. Information on the average UV photodesorption yield of astrophysically important ices exists for broadband UV lamp experiments. %(Oberg et al. 2007, 2009). 
UV fields around low-mass pre-main sequence stars, around shocks and in many other astrophysical environments are however often dominated by discrete atomic and molecular emission lines. %(Bergin et al. 2003). 
It is therefore crucial to consider the wavelength dependence of photodesorption yields and mechanisms.
In this work, for the first time, the wavelength-dependent photodesorption of pure CO ice is explored %in various spectral windows 
between 90 and 170 nm. The experiments are performed under ultra high vacuum conditions using tunable synchrotron radiation. Ice photodesorption is simultaneously probed by infrared absorption spectroscopy in reflection mode of the ice and by quadrupole mass spectrometry of the gas phase.
The experimental results for CO reveal a strong wavelength dependence directly linked to the vibronic transition strengths of CO ice, implying that photodesorption is induced by electronic transition (DIET). The observed dependence on the ice absorption spectra implies relatively low photodesorption yields at 121.6 nm (Lyman $\alpha$), where CO barely absorbs, compared to the high yields found at wavelengths coinciding with transitions into the first electronic state of CO (A$^1\Pi$ at 150 nm); the CO photodesorption rates %are thus 
depend %ent 
strongly on the UV profiles encountered in %the 
different star formation environments.
\end{abstract}


%%%%%%%%%%%%%%% Document %%%%%%%%%%%%%%%%%%

\section{Introduction}
%\textbf{Why is it interesting}
%\begin{itemize}

%Carbon monoxide is observed in the gas phase and in the form of ice condensed on the surface of sub-microns size dust grains in the interstellar medium.
Carbon monoxide, the second most abundant molecule observed in the gas phase of the interstellar medium (ISM) after H$_2$, is also one of the most commonly detected molecules in the solid phase, condensed on the surface of sub-micron sizes dust grains \citep[e.g.,][]{Pontoppidan_03}.
Mechanisms triggering CO phase transitions under densities and temperatures encountered in star forming environments are crucial to interpret observations of CO lines. In cold parts of the ISM, most of the CO is depleted onto the grains but non-thermal desorption mechanisms, such as UV photodesorption or cosmic ray sputtering, can maintain a part of the CO budget in %to 
the gas phase, explaining the presence of CO below its thermal desorption temperature \citep[e.g.,][]{Willacy_00}. %add ref to specific regions

Desorption of CO ice induced by UV photons under astrophysically relevant conditions has recently been studied by \cite{Oberg_07, Oberg_09}, and \cite{MunozCaro_10} using an H$_2$ based broadband microwave discharge lamp as photon source and monitoring the ice loss through infrared spectroscopy during irradiation. The derived photodesorption rates are substantially higher than previously assumed \citep{Greenberg_73}, but differ by up to a factor of 20 between the two groups. Absolute photodesorption yield values have substantial uncertainties resulting from (1) the different photon flux calibration -- \cite{Oberg_07, Oberg_09} used a NIST-calibrated photodiode whereas \cite{MunozCaro_10} employed a chemical actionometry method -- and (2) the IR measurement of the ice sublimation, performed in transmission in the case of \cite{MunozCaro_10}, and in reflection in the case of \cite{Oberg_07, Oberg_09}. Nonetheless, the combined uncertainties are estimated to a factor of a few and cannot fully account for the order of magnitude difference between the two groups. The different UV irradiation profiles of the two discharge lamps may instead be at the origin of this discrepancy if the photodesorption efficiency is wavelength dependent.

From these studies, excitation of CO in its first electronic state was proposed to induce desorption after energy transfer to neighboring CO molecules and rearrangement of the ice surface (M.C. van Hemert 2010, private communication). Wavelength-resolved studies are required to confirm this mechanism and to investigate whether the photodesorption mechanism of CO is wavelength dependent. 

Wavelength-resolved photodesorption yields are also important for astrochemical networks. Different FUV field profiles are encountered at different star formation stages and around young stellar objects of different spectral types. The FUV can be dominated by the interstellar radiation field (ISRF) at the edge of molecular clouds, cosmic-rays excited H$_2$ emission in starless clouds \citep{Gredel_87}, black body emission from protostars, and emission lines due to material accretion onto the protostar or pre-main-sequence star \citep{Bergin_03, vDishoeck_06}. Thus if photodesorption yields are wavelength-dependent, the photodesorption yield per incident UV photon may vary significantly in different star-forming environments. 

In order to fully characterize the CO photodesorption process, the present work investigates for the first time the wavelength dependence of CO photodesorption yields for astrophysically relevant ices and conditions. The experimental techniques for ice preparation and synchrotron-based irradiations are explained in Section  \ref{Sec_exp}. The results are described in Section \ref{Sec_res} and the mechanism and astrophysical implications are discussed in Section \ref{Sec_dis}.

 \section{Experimental}
\label{Sec_exp}

\begin{figure}[t]
  \centering
  \includegraphics[width=0.47\textwidth]{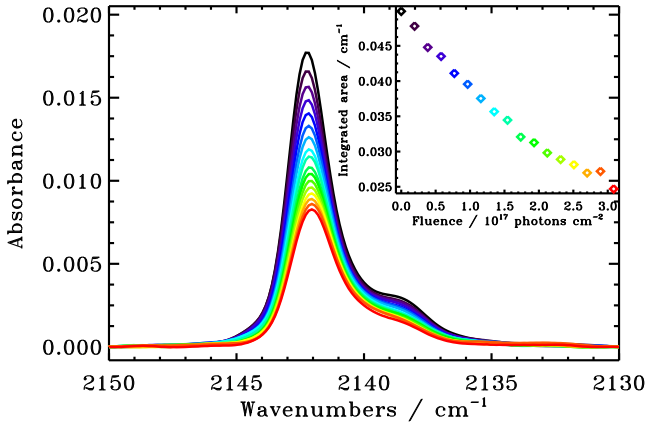}
  \caption{Decrease of the CO vibration RAIRS signal during irradiation of a CO ice 10 ML thick at 18 K by 9.2 eV photons. Inset shows the integrated area of the corresponding CO RAIRS band vs. UV fluence.  }
  \label{Fig_RAIRS}

\end{figure} 

CO ices are grown and irradiated in the SPICES (Surface Processes in ICES) apparatus, described in detail elsewhere \citep{Bertin_11}. CO ice films are prepared under ultra high vacuum (UHV) conditions (base pressure: $5\times10^{-10}$ Torr) on a polycrystalline gold substrate which is cooled down to $18~ \rm K$ by a closed cycle helium cryostat. The CO ices are grown via background deposition, i.e., by exposing the cold Au surface to a partial pressure of CO gas (Air Liquide, 99\% purity) during a given time. The amount of CO deposited on the surface is monitored through Reflection Absorption InfraRed Spectroscopy (RAIRS). The thickness of the deposited ice in monolayers (ML) is obtained by performing an isothermal desorption experiment of the CO ice monitored by RAIRS, resulting in the absorbance of 1 monolayer of CO ice \citep{Oberg_09}. Complementary estimations of the CO ice thickness have also been performed using Temperature Programmed Desorption (TPD) experiments and from the exposure, assuming a sticking coefficient of unity. All techniques gave similar results, yielding ice thicknesses between 9 and 10~ML.  

The set-up is equipped with two complementary photodesorption detection techniques. A Quadrupole Mass Spectrometer (QMS) allows the detection of gas phase molecules within the UHV chamber, while condensed molecules are probed by RAIRS. Simultaneous probing of the remaining condensed molecules and of the gas-phase photodesorbed molecules is therefore possible during ice irradiation. QMS measurements are used to determine the relative photodesorption rates for all employed UV fluxes, while RAIRS provides absolute yields for UV fluxes higher than 10$^{13}$ photons s$^{-1}$.

The UHV set-up is connected to the undulator-based FUV beamline DESIRS at the SOLEIL synchrotron facility in Saint-Aubin (France). In a first irradiation mode, the photon beam from the undulator stage is used directly. The high harmonics of the photon energy are filtered using a rare-gas filled chamber, set in the pass of the UV beam. Absolute incident photon fluxes are measured using calibrated photodiodes. This setting yields high photon flux (10$^{14}$ photons s$^{-1}$) at selected energies with a well defined photon energy distribution (Lorentzian with FWHM of 1 eV). Irradiations at fixed photon energies (8.2, 9.2, 10.2, and 11.2 eV) are performed to monitor the CO ice loss with photon fluence using RAIR spectroscopy (see Figure \ref{Fig_RAIRS}). These experiments are used to derive absolute photodesorption rates for different spectral windows where CO ice absorption is known to differ (see Section \ref{Sec:kinetics}). 

In a second irradiation mode, the photons enter a 6.65 m normal incidence monochromator equipped with a 200 grooves mm$^{-1}$ grating. In this configuration, a photon flux up to 10$^{12}$ photons s$^{-1}$ with a resolution of 40 meV at 10 eV for a 700 $\mu$m exit slit is achieved. The CO ice film is then irradiated by photons whose energy is linearly increased throughout the experiment, from 7 to 13.6 eV (90 to 170 nm). A relative photodesorption rate spectrum is obtained by recording the QMS signal of desorbing CO while tuning the photon energy (see upper panel of Figure \ref{Fig_spec} and Section \ref{Sec:spec}).

\section{Results}
\label{Sec_res}
\subsection{Absolute Photodesorption Yields for 1~eV Spectral Windows}
\label{Sec:kinetics}

\begin{deluxetable}{l c c}
\tablecaption{Energy dependent photodesorption rates for 10~ML thick CO ice at 18~K.}
\label{Tab_rates} 
\tablehead{ \colhead{Irradiation Energy} & \colhead{Relative Rate} & \colhead{Absolute Rate} \\ \colhead{(eV)} & \colhead{(cm$^{-1}$ (photons cm$^{-2}$)$^{-1}$)} & \colhead{(molecules photon$^{-1}$)}}
\startdata
8.2 & $2.19 \times 10^{-19}$ & $2.8 \pm 1.7 \times 10^{-2}$ \\
 9.2 &  $1.07 \times 10^{-19}$& $1.3 \pm 0.91 \times 10^{-2}$ \\
 10.2 & $0.54 \times 10^{-19}$& $6.9 \pm 2.4 \times 10^{-3}$ \\
11.2 & $0.74 \times 10^{-19}$& $9.3 \pm 3.4 \times 10^{-3}$ 
\enddata
\end{deluxetable}

Figure \ref{Fig_RAIRS} shows the decrease of the CO stretching feature during irradiation of a 10 ML thick ice at 18 K by 9.2 eV photons at a flux of $1.2 \times 10^{14} \rm photon s^{-1}$.  The double-peaked structure of the infrared feature does not change during ice growth and is most likely due to the roughness of the gold substrate; the peak structure is known to be sensitive to the arrangement of adsorbed molecules on the surface \citep{Palumbo_06}.
The evolution of the CO absorption band integrated area as a function of photon fluence is shown in the inset. The integrated band area is directly proportional to the amount of CO ice (from RAIR spectra acquired during ice growth). The loss of integrated band area during irradiation can therefore be used to  quantify CO  desorption. 

The observed linear relationship between CO ice integrated area and photon fluence is consistent with a zeroth-order process, similar to what has been seen for broadband irradiation experiments \citep{Oberg_07, MunozCaro_10}. This zeroth-order kinetics is observed for the four explored irradiation energies (8.2, 9.2, 10.2, and 11.2 eV) until the ice coverage gets below 0.03 cm$^{-1}$ integrated absorbance ($\sim$6 ML). The same effect has been observed by \cite{MunozCaro_10}. Fitting straight lines to the linear parts of the ice loss at the four investigated irradiation energies provides energy-dependent photodesorption yields.

The results from the fits are summarized in Table \ref{Tab_rates}. The relative photodesorption yields (absorbance per incident photon flux) are converted into absolute photodesorption yields (CO molecules per incident photon)  using the absorbance to ML conversion factor for the present RAIRS set-up (Section \ref{Sec_exp}) and assuming that 1 ML = 10$^{15}$ molecules cm$^{-2}$. The uncertainties in the relative yields are due to the intrinsic RAIRS measurement uncertainties. The larger uncertainties on the absolute yields are due to ice thickness calibration and to differences in the FUV irradiated area versus IR probed area. Indeed, the sample area probed by the IR beam and the UV-beam spot at the surface are $1 \pm 0.1\, \rm cm^2$ and $0.7\pm 0.1\, \rm cm^2$, respectively. Caution has been taken to ensure that the UV irradiated area is integrally probed by the IR beam. The photodesorption yields vary between $\rm 6.9 \, and \,  28 \times 10^{-3}$ molecules photon$^{-1}$ for the investigated energy range and are thus clearly wavelength-dependent. The lowest yield is obtained at 10.2 eV, which is important considering that previous photodesorption results were based on broadband UV discharge lamps, which are often dominated by photons at this excitation energy. In addition, the strong variation with photon energy motivates a more detailed investigation on the energy dependence of the photodesorption yield.

\subsection{CO Photodesorption Yield Spectrum}
\label{Sec:spec}

	 \begin{figure}[t]
  \centering
   \includegraphics[width=0.47\textwidth]{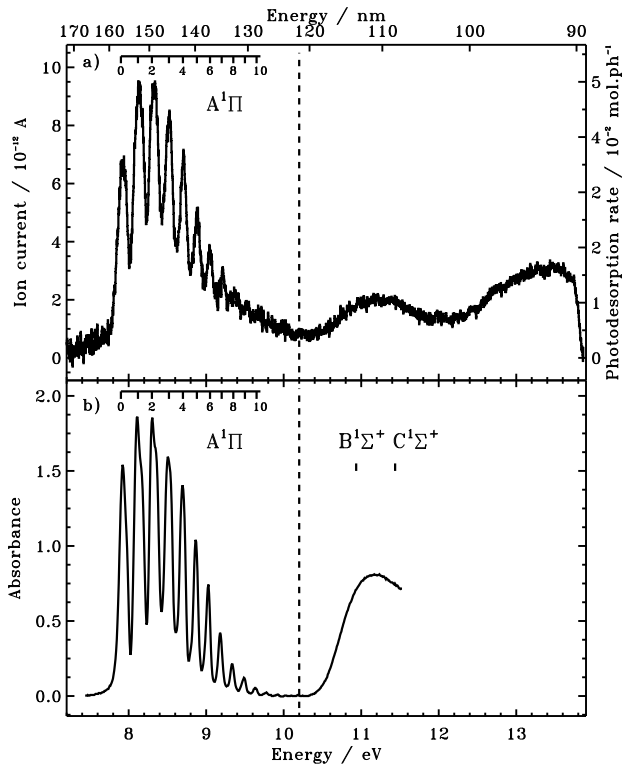}
      \caption{(a) Photodesorption spectrum (after background subtraction) of a 10 ML CO ice at 18 K using a resolution of 40 meV (at 10 eV) and a scan step of 12 meV every 0.5 s.  (b) Absorption spectrum of solid CO at 10 K \citep{Lu_05}. The dashed line indicates photodesorption and absorption values at Ly$\alpha$, 121.6 nm.}
              
         \label{Fig_spec}
   \end{figure} 

A detailed investigation of the wavelength dependence is achieved in the second kind of experiment, where a 10~ML thick CO ice sample is irradiated by a monochromatic photon beam scanning 7--13.6~eV at a constant rate, while probing photodesorption with the QMS. The QMS current resulting from the desorbing gas phase CO ($m/z = 28$) versus photon energy is presented in Figure \ref{Fig_spec} together with a UV absorption spectra of CO ice from \cite{Lu_05}. This photodesorption spectrum is converted into molecules photon$^{-1}$ using the absolute photodesorption yields, obtained from the first series of experiments (Section \ref{Sec:kinetics}, Table \ref{Tab_rates}). The scaling takes into account the exact spectral profiles in the 1~eV spectral-window experiments, which was monitored carefully for the 8.2~eV window. Figure \ref{Fig_spec}(a) shows that when probed at a high spectral resolution, the photodesorption yield varies even more across the investigated energy range compared to the 1~eV spectral-window measurements. The lowest yield is achieved below 7.8~eV and at 10.2~eV ($<6\times10^{-3}$ molecules photon$^{-1}$) and it is an order of magnitude lower than the peak value of $5\times10^{-2}$ molecules photon$^{-1}$ at $\sim$~8.2~eV. The average UV profile below 10 eV results in a total photodesorption yield of $1.8 \times 10^{-2} \rm molecules\, photon^{-1}$. This value is a factor of seven higher than the previously reported yield of $2.7 \times 10^{-3} \rm molecules\, photon^{-1}$ from \cite{Oberg_09}, and comparable within a factor of two to the yield of $3.5 \times 10^{-2} \rm molecules\, photon^{-1}$ found by \cite{MunozCaro_10}. The relative contribution of Ly$\alpha$ photons (and their inefficiency at inducing photodesorption) to the total photon flux in each experiment is a probable origin for the different yields, further stressing the importance of wavelength resolved studies.

\section{Discussion}
\label{Sec_dis}
\subsection{Mechanism}

The photodesorption spectrum (Figure \ref{Fig_spec}(a)) has a band structure below 10 eV that is almost identical to the absorption spectrum of pure CO ice obtained by \cite{Lu_05}. The observed spectral progression is attributed to the transition from CO ground state to vibrational levels of the first allowed electronic state ($\rm A ^1\Pi$). The striking similarity between the two spectra provides the first experimental evidence that desorption is a DIET (Desorption Induced by Electronic Transition) process, as previously suggested by \cite{Oberg_07, Oberg_09}. Desorption of condensed molecules is probably induced by subsequent relaxation of the excited molecules via energetic transfer from electronic to vibrational degrees of freedom \citep{Avouris_89}.
 This process clearly dominates the spectrum below 10 eV. Above 10~eV, the main photodesorption mechanism is expected to be more complex since the second absorption band of CO is dissociative. In this case, photodesorption may result from several relaxation pathways, including chemical recombination and desorption, as proposed in the case of water \citep{Westley_95, Watanabe_00, Arasa_10}.

It should be noted that electrons, resulting from the interaction of incident UV photons with the metallic substrate, may also contribute to the observed CO ice desorption. Most of the UV photons can be absorbed by the metallic substrate since the ice coverage is an order of magnitude below the  coverage limit for which most of the photons would be absorbed by the ice before reaching the substrate. Hot electrons resulting from electron-hole pair excitation of gold are expected to be produced for all investigated irradiation energies and may induce desorption (e.g., \cite{Bonn_99}). Free electrons photo-emitted from the gold substrate may also induce CO desorption by electronic excitation or resonant electron attachment \citep{Schultz_73, Rakhovskaia_95, Shi_98, Mann_95}.  Only photons whose energy is above 10 eV can produce secondary free electrons energetic enough to trigger the latter process, since the CO-covered gold surface work function is $\sim 4.4 \rm eV$ \citep{Gottfried_03}. The importance of these electron-induced processes can be estimated from our experiment, since the production of electrons (primary photoelectrons and ineslatically diffused secondary electrons) that have enough kinetic energy to trigger the CO desorption process within the ice, should increase with the photon energy. This effect is barely observed; electrons are most likely the cause of the low-level continuously increasing desorption background seen in Figure \ref{Fig_spec}(a). In other words, DIET is the main pathway leading to the observed photodesorption below 10 eV, resulting in a strong energy dependence of the photodesorption yields.

\subsection{Astrophysical implications}

The determined strong wavelength dependence of the photodesorption yield of CO ice presented in Figure \ref{Fig_spec}(a) demonstrates that the CO photodesorption yield is expected to vary significantly in different astrophysical environments. %should be used to improve %spectrum provides new data to improve 
Incorporating this information into astrochemical networks is important to accurately predict the partitioning of CO between the gas and ice. %which includ UV-dependent chemistry. 
Table \ref{Tab_env} exemplifies how this wavelength dependent data can be used to predict the photodesorption efficiency %more accurately 
in sources with characteristic UV fields. The listed photodesorption yields have been calculated by convolving UV field profiles %for different environments have been collected and convoluted 
from the literature with the CO photodesorption spectrum presented in Figure \ref{Fig_spec}(a) between 90 and 180 nm. % to yield characteristic photodesorption. 
The UV profiles %used here 
are the ISRF from \cite{Mathis_83} appropriate to model %to illustrate 
the UV field at the edges of molecular clouds, an emission spectrum of cosmic-ray-excited H$_2$ \citep{Gredel_87} to mimic UV field encountered in %re-stellar 
cloud cores, a 10,000~K black body for the UV field around %coming from 
an Herbig Ae star \citep{vDishoeck_06} and the emission of TW Hydrae from \cite{Bergin_03} to illustrate photodesorption in protoplanetary disks. 
Pure Ly$\alpha$ is also added for comparison. The calculated characteristic photodesorption yields vary by a factor of four, with the lowest yields around T Tauri stars and other environments where the UV field is dominated by Ly$\alpha$ photons, and the highest yields per incident photon at cloud edges and next to a 10,000~K black body. The photodesorption yields used here are valid for pure CO ice thicker than $\sim 6\, \rm ML$, as stated in Section \ref{Sec:kinetics}. How the CO photodesorption efficiency and mechanism change with ice thickness and/or composition should certainly be studied in order to extend the applicability of the present results to thinner pure CO ices or mixed CO containing astrophysical ices.

%From this table, it can be seen that, because the UV distribution for the T-Tauri environment is peaked at Lyman-alpha, CO photodesorption efficiency around TW Hydrae is lower, by almost a factor 2, than at the edges of molecular clouds.

%{\bf[add comments about the other environments as well to guide the reader. Be specific and quantitative.]} 

It is important to note %One should acknowledge 
that the UV spectral profile deeper into the cloud and disk may be radically different compared to the UV field incident on the cloud and disk surface because of wavelength-dependent radiative transfer. Determining accurate interstellar relevant photodesorption yields requires a combination of  photodesorption spectrum and detailed UV radiative transfer. In the meantime, the presented photodesorption spectrum is recommended to calculate photodesorption yields for specific environments (Table \ref{Tab_env}).
  \begin{deluxetable}{l c c}
  
\tablecaption{CO Photodesorption in Different ISM Environments}             % title of Table
\label{Tab_env} 
%\begin{tabular}{l l l}   
   
\tablehead{ \colhead{Environment} & \colhead{Yield} & \colhead{References for UV profile} \\ \colhead{ } &  \colhead{(molecules photon$^{-1}$)} & \colhead{ } }
             % inserts double horizontal lines
%Environment & Yield& References for UV profile\\
% &\hspace{-0.3 cm}\begin{scriptsize}molecules photon$^{-1}$\end{scriptsize}& \\
%\hline
\startdata
Edges of clouds & $1.2 \times 10^{-2}$  & \cite{Mathis_83} \\
Pre-stellar cores & $9.4 \times 10^{-3}$& \cite{Gredel_87} \\
Black body 10,000K & $ 1.6 \times 10^{-2}$ & \cite{vDishoeck_06}\\
Tw~Hydrae & $6.6 \times 10^{-3}$ & \cite{Bergin_03} \\
Pure ly$\alpha$ & $4.1 \times 10^{-3}$& 121.6 nm
\enddata
\end{deluxetable}

% according to Fig. \ref{Fig_spec} a).

%\underline{Extra paragraph with info }: Calculation made for UV unattenuated by dust .the UV profile is changed with dust scattering and H2 see Bergin 11 paper. Also we don't have a pure CO ice so this may change once we have more detailed ice - grain model which take into account the fact that species are mixed but considering that we are looking at the CO-rich mantle, approximation here is reasonable and should be used until more elaborated ice code are made available.

 %\textbf{ Warning : for the second absorption band we have desorbtion subsquent to dissociation and that in the case of a real instellar ice the other species present in the CO layer may react with the fragment so we expect the photodesorption rate to be different at those energies and be a function of the other species concentrations. }
 
 \section{Conclusions}

This study provides the first photodesorption spectrum of pure CO ice, obtained by tunable synchrotron UV irradiation of CO ice under astrophysically relevant conditions. A quantitative determination of the photodesorption yields at 8.2--13.6~eV has been achieved by the simultaneous probe of the ice- and gas- phase CO concentrations coupled to either narrowband (1~eV) excitation mode at selected energies or to continuous energy scanning by monochromatic UV radiation. The resulting photodesorption yields vary by an order of magnitude over the investigated wavelength range. The CO photodesorption process is dominated by the direct electronic excitation of the condensed CO molecules (DIET) below 10~eV. Other desorption mechanisms involving secondary electrons due to UV interactions with the substrate are present, but their contributions to the total measured desorption yield are minor. Consistent with a DIET mechanism, the photodesorption yield of CO ice at Ly$\alpha$ is low. This may result in different chemical evolutions in regions where the UV field is dominated by line emission compared to regions where the UV emission is due to black body radiation. The calibrated photodesorption spectrum presented here should therefore be used to determine the characteristic CO ice photodesorption yields by convolving the CO photodesorption spectrum to the UV profile found in a specific astrophysical environment. In general, the link between wavelength-dependent UV absorption of the ice and the resulting photodesorption and photochemistry is key to implement astrochemical gas-grain models and should be experimentally investigated for relevant ice species.

\acknowledgments
We are grateful to Ewine van Dishoeck and Marc van Hemert for stimulating discussions. % and comments on this manuscript. 
We thank Hsiao-Chi Lu, Roland Gredel, and Edwin Bergin for providing useful data. We acknowledge SOLEIL for provision of synchrotron radiation facilities, as well as Laurent Nahon for technical help on the beamline DESIRS. Financial supports from the French national program PCMI (Physique chimie du milieu interstellaire), the Dutch program NOVA (Nederlandse Onderzoekschool Voor Astronomie), the COST action CM0805 "The Chemical Cosmos", and the Hubert Curien Partenership "Van Gogh", are gratefully acknowledged. Support for K.I.\"O is provided by NASA through Hubble Fellowship grant awarded by the Space Telescope Science Institute, which is operated by the Association of Universities for Research in Astronomy, Inc., for NASA, under contract NAS 5-26555.

\bibliographystyle{apj}

\end{document}